\documentclass[aps,twocolumn,prl]{revtex4}
\usepackage{graphicx}
\usepackage{subfigure}

\begin{document}

\title{Theory of Ferroelectric Nanoparticles in Nematic Liquid Crystals}

\author{Lena M. Lopatina}
\affiliation{Liquid Crystal Institute, Kent State University, Kent, OH 44242}
\author{Jonathan V. Selinger}
\email{jvs@lci.kent.edu}
\affiliation{Liquid Crystal Institute, Kent State University, Kent, OH 44242}

\date{February 18, 2009}

\begin{abstract}

Recent experiments have reported that ferroelectric nanoparticles have drastic
effects on nematic liquid crystals---increasing the isotropic-nematic
transition temperature by about 5~K, and greatly increasing the sensitivity to
applied electric fields. To understand these effects, we develop a theory for
the statistical mechanics of ferroelectric nanoparticles in liquid crystals.
This theory predicts the enhancements of liquid-crystal properties, in good
agreement with experiments. These predictions apply even when electrostatic
interactions are partially screened by moderate concentrations of ions.

\end{abstract}
\maketitle

In recent years, many experiments have found that colloidal particles in
nematic liquid crystals exhibit remarkable new types of physical phenomena. If
the particles are micron-scale, they induce an elastic distortion of the
liquid-crystal director.  This elastic distortion leads to an effective
interaction between particles, and offers the possibility of organizing a
periodic array of particles, with possible photonic
applications~\cite{poulin97,gu00,stark01,yada04,smalyukh05,musevic06}.  If the
particles are 10--100 nm in diameter, they are too small to distort the
liquid-crystal director, and hence the system enters another range of
behavior.  Experiments have shown that low concentrations of
\textit{ferroelectric} nanoparticles can greatly enhance the physical
properties of nematic liquid crystals~\cite{reznikov03,ouskova03,%
reshetnyak04,buchnev04,glushchenko06,reshetnyak06,li06,buchnev07}.  In
particular, Sn$_2$P$_2$S$_6$ or BaTiO$_3$ nanoparticles at low concentration
($<$1\%) increase the orientational order parameter of the host liquid
crystal, and increase the isotropic-nematic transition temperature by about
5~K.  The nanoparticles also decrease the switching voltage for the
Fredericksz transition.  These experimental results are important for
fundamental nanoscience, because they show that nanoparticles can couple to
the orientational order of a macroscopic medium.  They are also important for
technological applications, because they provide a new opportunity to tune the
properties of liquid crystals without additional chemical synthesis.

A key question in this field is how to understand and control the properties
of liquid crystals doped with ferroelectric nanoparticles.  One paper has
argued that the nanoparticles produce large local electric fields, which
polarize the liquid-crystal molecules and thereby increase the intermolecular
interaction~\cite{li06}.  This increased interaction leads to a higher
isotropic-nematic transition temperature $T_\mathrm{NI}$.  A limitation of
this approach is that it does not consider the orientational distribution of
the nanoparticles themselves.  One might expect this distribution to be
crucial for understanding how nanoparticles affect the orientational order of
the liquid crystal.

In this paper, we propose a new theory for the statistical mechanics of
ferroelectric nanoparticles in liquid crystals, which is based specifically on
the orientational distribution of the nanoparticle dipole moments.  This
distribution is characterized by an orientational order parameter, which
interacts with the orientational order of the liquid crystals and stabilizes
the nematic phase.  We estimate the coupling strength and calculate the
resulting enhancement in $T_\mathrm{NI}$, in good agreement with experiments.
This enhancement occurs even when electrostatic interactions are partially
screened by moderate concentrations of ions in the liquid crystal.  In
addition, we predict the response of the isotropic phase to an applied
electric field, known as the Kerr effect, and show that it is greatly enhanced
by the presence of nanoparticles.

\begin{figure}[b]
(a)\subfigure{\includegraphics[scale=0.636,clip=true]{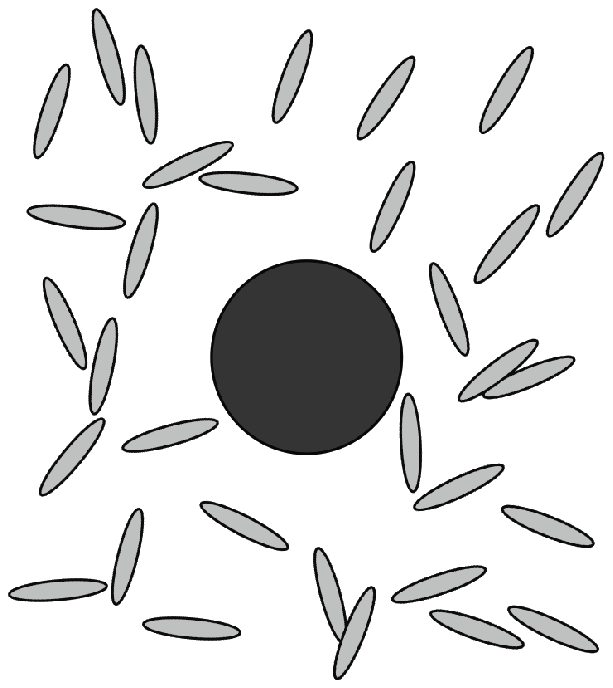}}
(b)\subfigure{\includegraphics[scale=0.636,clip=true]{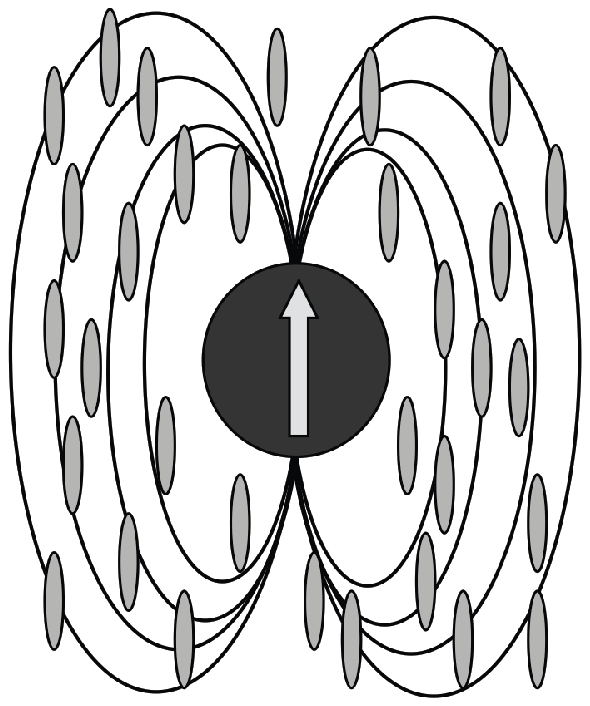}}
\caption{Nanoparticles surrounded by liquid crystal.  (a) Particle with no
electric dipole moment, in the isotropic phase.  (b) Ferroelectric particle
with electric dipole moment, which produces an electric field that interacts
with the orientational order of the liquid crystal.}
\end{figure}

To begin the calculation, consider a spherical nanoparticle with radius $R$
and electrostatic dipole moment $\mathbf{p}$, surrounded by a nematic liquid
crystal, as shown in Fig.~1.  The electric field $\mathbf{E}$ generated by the
nanoparticle interacts with the order tensor $Q^\mathrm{LC}_{\alpha\beta}$ of
the liquid crystal through the free energy
\begin{equation}
F_\mathrm{int}=-\frac{\epsilon_0\Delta\epsilon}{3} \int d^3 r
Q^\mathrm{LC}_{\alpha\beta}(\mathbf{r})
E_\alpha(\mathbf{r}) E_\beta(\mathbf{r}),
\end{equation}
where $\Delta\epsilon$ is the dielectric anisotropy of the fully aligned
liquid crystal.  The electric field of the nanoparticle has the standard
dipolar form
\begin{equation}
\mathbf{E}(\mathbf{r})=\frac{1}{4\pi\epsilon_0\epsilon}\left(
\frac{3\mathbf{r}(\mathbf{r}\cdot\mathbf{p})}{r^5}-\frac{\mathbf{p}}{r^3}
\right),
\end{equation}
neglecting higher-order corrections due to the dielectric anisotropy of the
liquid crystal.  Near the nanoparticle, the electric field varies rapidly as a
function of position.  However, the liquid-crystal order cannot follow that
rapid variation, because it would cost too much elastic energy.  Hence, for
sufficiently small nanoparticles, we can assume that order tensor
$Q^\mathrm{LC}_{\alpha\beta}$ is uniform in space.  In that case, we can
integrate the interaction free energy to obtain
\begin{equation}
F_\mathrm{int}=-\frac{\Delta\epsilon}{180\pi\epsilon_0\epsilon^2 R^3}
Q^\mathrm{LC}_{\alpha\beta} p_\alpha p_\beta.
\end{equation}

Now consider a low concentration $\rho_\mathrm{NP}$ of nanoparticles dispersed
in the liquid crystal.  The dipole moments of these nanoparticles will not all
have the same orientation; rather there must be a distribution of
orientations.  As a result, the interaction of liquid crystal and
nanoparticles gives the free energy density per unit volume
\begin{equation}
\frac{F_\mathrm{int}}{V}=
-\frac{\Delta\epsilon \rho_\mathrm{NP}}{180\pi\epsilon_0\epsilon^2 R^3}
Q^\mathrm{LC}_{\alpha\beta} \langle p_\alpha p_\beta \rangle,
\end{equation}
averaged over the distribution of nanoparticle orientations.  This
distribution can be expressed in terms of a nanoparticle order tensor
\begin{equation}
Q^\mathrm{NP}_{\alpha\beta}=\frac{3}{2}
\frac{\langle p_\alpha p_\beta \rangle}{p^2}-\frac{1}{2}\delta_{\alpha\beta},
\end{equation}
analogous to the standard liquid-crystal order tensor.  Hence, the
interaction free energy density becomes
\begin{equation}
\frac{F_\mathrm{int}}{V}=
-\frac{\Delta\epsilon \rho_\mathrm{NP} p^2}{270\pi\epsilon_0\epsilon^2 R^3}
Q^\mathrm{LC}_{\alpha\beta} Q^\mathrm{NP}_{\alpha\beta},
\end{equation}
which shows an explicit coupling between the order tensor of the liquid
crystal and the order tensor of the nanoparticles.  If we make the reasonable
assumption that both of these tensors are aligned along the same axis, then
this interaction reduces to
\begin{equation}
\frac{F_\mathrm{int}}{V}=
-\frac{\Delta\epsilon \rho_\mathrm{NP} p^2}{180\pi\epsilon_0\epsilon^2 R^3}
S_\mathrm{LC} S_\mathrm{NP},
\end{equation}
where $S_\mathrm{LC}$ and $S_\mathrm{NP}$ are the scalar order parameters of
the liquid crystal and the nanoparticles, respectively.

To model the statistical mechanics of nanoparticles dispersed in the liquid
crystal, we must expand the free energy in \emph{both} order parameters
$S_\mathrm{LC}$ and $S_\mathrm{NP}$, which gives
\begin{eqnarray}
\frac{F}{V}&=&\displaystyle{\frac{a'_\mathrm{LC}(T-T^*)}{2}S_\mathrm{LC}^2
-\frac{b}{3}S_\mathrm{LC}^3
+\frac{c}{4}S_\mathrm{LC}^4} \nonumber\\
&&\displaystyle{+\frac{a_\mathrm{NP}}{2}S_\mathrm{NP}^2
-\frac{\Delta\epsilon \rho_\mathrm{NP} p^2}{180\pi\epsilon_0\epsilon^2 R^3}
S_\mathrm{LC} S_\mathrm{NP}.}
\label{freeenergy}
\end{eqnarray}
Here, the first three terms are the standard Landau-de Gennes free energy of a
nematic liquid crystal.  The first-order isotropic-nematic transition of the
pure liquid crystal occurs at
$T_\mathrm{NI}=T^* + (2 b^2)/(9 a'_\mathrm{LC} c)$, and the leading
coefficient in this expansion can be estimated through Maier-Saupe theory as
$a'_\mathrm{LC}=5 k_\mathrm{B} \rho_\mathrm{LC}$, where $\rho_\mathrm{LC}$ is
the concentration of liquid-crystal molecules per volume~\cite{katriel86}.
The fourth term in the free energy is the entropic cost of imposing
orientational order on the nanoparticles.  By expanding the entropy in terms
of the orientational distribution function of the nanoparticles, we can
estimate the coefficient as $a_\mathrm{NP}=5k_\mathrm{B} T \rho_\mathrm{NP}$.
The final term is the coupling between the liquid-crystal order and the
nanoparticle order, calculated above.

We minimize the free energy of Eq.~(\ref{freeenergy}) over the nanoparticle
order parameter to find the optimum value
\begin{equation}
S_\mathrm{NP}=
\frac{\Delta\epsilon p^2}{900\pi\epsilon_0\epsilon^2 R^3 k_\mathrm{B} T}
S_\mathrm{LC}.
\end{equation}
This equation shows that the liquid crystal induces orientational order of the
nanoparticles, with a nanoparticle order parameter proportional to the
liquid-crystal order parameter.  Note that the induced order is independent of
the nanoparticle concentration, which is reasonable because it arises from
interaction of individual nanoparticles with the liquid crystal, not from
interactions between nanoparticles.  We then substitute this expression back
into the free energy to obtain
\begin{eqnarray}
\frac{F}{V}&=&\displaystyle{\frac{a'_\mathrm{LC}}{2}
\left[T-T^*-\frac{\rho_\mathrm{NP}^2}{a'_\mathrm{LC}a_\mathrm{NP}}
\left(\frac{\Delta\epsilon p^2}{180\pi\epsilon_0\epsilon^2 R^3}\right)^2
\right]S_\mathrm{LC}^2} \nonumber\\
&&\displaystyle{-\frac{b}{3}S_\mathrm{LC}^3 +\frac{c}{4}S_\mathrm{LC}^4 .}
\end{eqnarray}
In this equation, the coefficient of $S_\mathrm{LC}^2$ has been shifted by the
interaction with the nanoparticles.  This shift increases the
isotropic-nematic transition temperature by
\begin{eqnarray}
\Delta T_\mathrm{NI}&=&
\displaystyle{\frac{\rho_\mathrm{NP}^2}{a'_\mathrm{LC}a_\mathrm{NP}}
\left(\frac{\Delta\epsilon p^2}{180\pi\epsilon_0\epsilon^2 R^3}\right)^2}
\nonumber\\
&=&
\displaystyle{\frac{\pi\phi_\mathrm{NP}R^3}{3T_\mathrm{NI}\rho_\mathrm{LC}}
\left(\frac{2\Delta\epsilon P^2}{675k_\mathrm{B} \epsilon_0\epsilon^2}
\right)^2} .
\label{DeltaTNI}
\end{eqnarray}
The last expression has been simplified by writing $p=(\frac{4}{3}\pi R^3)P$
and $\rho_\mathrm{NP}=\phi_\mathrm{NP}/(\frac{4}{3}\pi R^3)$, where $P$ is the
polarization and $\phi_\mathrm{NP}$ the volume fraction of the nanoparticles.

To estimate $\Delta T_\mathrm{NI}$ numerically, we use the following
parameters appropriate for Sn$_2$P$_2$S$_6$ nanoparticles in the liquid
crystal 5CB:
$\phi_\mathrm{NP}=0.5\%$,
$R=35$ nm,
$T_\mathrm{NI}=308$ K,
$\rho_\mathrm{LC}=2.4\times10^{27}$ m$^{-3}$,
$P=0.04$ Cm$^{-2}$,
$k_\mathrm{B}=1.38\times10^{-23}$ JK$^{-1}$,
$\epsilon_0=8.85\times10^{-12}$ C$^2$N$^{-1}$m$^{-2}$, and
$\Delta\epsilon\approx\epsilon\approx 10$~\cite{reznikov}.  With those
parameters, we obtain $\Delta T_\mathrm{NI}\approx 5$~K, which is roughly
consistent with the increase that is observed experimentally.  Of course,
there is a substantial uncertainty in this estimate, because the parameters
$R$ and $P$ are not known very precisely in the experiments.

Note that our model predicts that the enhancement $\Delta T_\mathrm{NI}$
should be first-order in volume fraction $\phi_\mathrm{NP}$, fourth-order in
polarization $P$, and third-order in $R$.  In particular, increasing $R$
should increase $\Delta T_\mathrm{NI}$ as long as the nanoparticles are not
large enough to disrupt the liquid-crystal order.  This prediction disagrees
with Ref.~\cite{li06}, which predicts that $\Delta T_\mathrm{NI}$ should be
first-order in $\phi_\mathrm{NP}$, second-order in $P$, and independent of
$R$.

At this point, we must consider the effects of ionic impurities in the liquid
crystal.  Any liquid crystal contains some concentration of free positive and
negative ions, which can redistribute in response to electric fields.  One
might worry that these ions would screen the electric field of the
nanoparticles, and hence prevent the enhancement of $T_\mathrm{NI}$.  To
address this issue, we solve the linearized Poisson-Boltzmann equation to find
the electric field around a dipole in the presence of ions
\begin{equation}
\mathbf{E}(\mathbf{r})=\frac{e^{-\kappa r}}{4\pi\epsilon_0\epsilon}
\left[(1+\kappa r)
\left(\frac{3\mathbf{r}(\mathbf{r}\cdot\mathbf{p})}{r^5}
-\frac{\mathbf{p}}{r^3}\right)
+\frac{\kappa^2 \mathbf{r}(\mathbf{r}\cdot\mathbf{p})}{r^3}\right].
\end{equation}
In this expression, $\kappa^{-1}$ is the Debye screening length given by
\begin{equation}
\kappa^{-1}=
\left(\frac{\epsilon_0\epsilon k_\mathrm{B} T}{2 n q^2}\right)^{1/2},
\end{equation}
where $n$ is the concentration and $q$ the charge of the ions.  With this
expression for the field, we can repeat the calculation above for the
enhancement in $T_\mathrm{NI}$, leading to
\begin{eqnarray}
\Delta T_\mathrm{NI}
&=&\displaystyle{\frac{\pi\phi_\mathrm{NP}R^3}{3T_\mathrm{NI}\rho_\mathrm{LC}}
\left(\frac{2\Delta\epsilon P^2}{675k_\mathrm{B} \epsilon_0\epsilon^2}
\right)^2 e^{-2\kappa R}}\nonumber\\
&&\displaystyle{\times\left(1 +2\kappa R +\kappa^2 R^2 +\kappa^3 R^3 \right).}
\end{eqnarray}

To interpret this result, note that the key parameter is $\kappa R$, the ratio
of the nanoparticle radius to the Debye screening length.  If the ion
concentration is low, then the screening length is large compared with the
nanoparticle radius, and hence $\Delta T_\mathrm{NI}$ is as large as in the
unscreened case.  However, if the ion concentration is sufficiently high,
then the screening length becomes comparable to the nanoparticle radius, and
hence the enhancement is screened away.  For a specific example, Fig.~2 shows
$T_\mathrm{NI}$ as a function of ion concentration $n$, using the numerical
parameters discussed above.  In this example, the full unscreened enhancement
persists up to $n\approx 10^{20}$ ions/m$^3$.  It then decays away as a
function of ion concentration, and is virtually eliminated by
$n\approx 10^{23}$ ions/m$^3$.

Typical measurements of the ion concentration
in 5CB show $n\approx 10^{20}$ ions/m$^3$ and hence $\kappa^{-1}\approx 260$
nm~\cite{sawada98}.  Because this screening length is much greater than the
nanoparticle radius, the enhancement should indeed be observable in realistic
experiments.  Note that the ion concentration in a liquid crystal varies over
several orders of magnitude, depending on preparation conditions.  Hence, we
speculate that variations in ion concentration may be one explanation for
variations in published experimental measurements of $\Delta T_\mathrm{NI}$.

\begin{figure}
\includegraphics[width=3.375in]{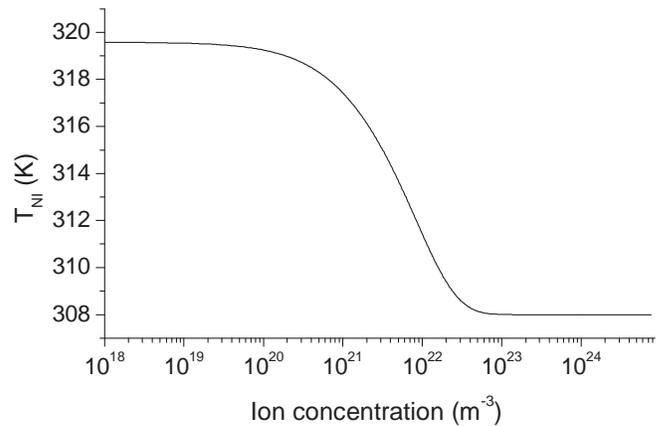}
\caption{Predicted isotropic-nematic transition temperature as a function of
ion concentration in a nanoparticle-doped liquid crystal, using numerical
parameters presented after Eq.~(\ref{DeltaTNI}).}
\end{figure}

So far, we have modeled the spontaneous ordering of a nanoparticle-doped
liquid crystal.  We can also use the same theoretical approach to predict how
the system responds to an applied electric field.  For a specific example, we
investigate the Kerr effect, in which an applied electric field $\mathbf{E}$
induces orientational order in the isotropic phase, slightly above the
isotropic-nematic transition.  In a pure liquid crystal, the Kerr effect is a
weak alignment proportional to $E^2$.  In a liquid crystal doped with
ferroelectric nanoparticles, we expect that an applied electric field will
induce \textit{polar} order of the nanoparticles, proportional to $E$.  This
polar order will necessarily induce \textit{nematic} order of the
nanoparticles, proportional to $E^2$, which will in turn induce nematic order
of the liquid crystal, also proportional to $E^2$.  Hence, the
nanoparticle-doped liquid crystal should have an enhanced Kerr effect with the
same symmetry as the standard Kerr effect, but with a much larger magnitude.

To model the enhanced Kerr effect, we must generalize the Landau theory
presented above in three ways.  First, we must introduce a polar order
parameter $M_\alpha=\langle p_\alpha\rangle/p$ for the nanoparticles, as well
as the nematic order parameters $Q^\mathrm{NP}_{\alpha\beta}$ and
$Q^\mathrm{LC}_{\alpha\beta}$.  Second, we must consider the energetic
coupling of an applied electric field to the order parameters.  In the free
energy density, an applied field couples linearly to the polar order parameter
of the nanoparticles through the interaction
$-\rho_\mathrm{NP} p E_\alpha M_\alpha$, and couples quadratically to the
nematic order parameter of the liquid crystal through the interaction
$-\frac{1}{3}\epsilon_0 \Delta\epsilon E_\alpha E_\beta
Q^\mathrm{LC}_{\alpha\beta}$.  Third, we must calculate the entropy of a
nanoparticle distribution characterized by both order parameters $M_\alpha$
and $Q^\mathrm{NP}_{\alpha\beta}$, following the method of
Ref.~\cite{katriel86}.  Assuming that all order parameters are aligned along
the electric field direction, the free energy becomes
\begin{eqnarray}
\frac{F}{V} &=& \displaystyle{\frac{a'_\mathrm{LC}(T-T^*)}{2}S_\mathrm{LC}^2
-\frac{b}{3}S_\mathrm{LC}^3
+\frac{c}{4}S_\mathrm{LC}^4}
-\frac{\epsilon_0\Delta\epsilon}{3}E^2 S_\mathrm{LC} \nonumber\\
&&\displaystyle{
-\frac{\Delta\epsilon \rho_\mathrm{NP} p^2}{180\pi\epsilon_0\epsilon^2 R^3}
S_\mathrm{LC} S_\mathrm{NP} } -\rho_\mathrm{NP} p E M\\
&&\displaystyle{+k_\mathrm{B}T\rho_\mathrm{NP}
\left(\frac{5}{2}S_\mathrm{NP}^2 +\frac{3}{2}M^2
-3 S_\mathrm{NP} M^2 \right). \nonumber}
\end{eqnarray}
We minimize this free energy over all three order parameters, $M$,
$S_\mathrm{NP}$, and $S_\mathrm{LC}$.  In the high-temperature isotropic
phase, in the limit of small electric field, the resulting liquid-crystal
order parameter is
\begin{equation}
S_\mathrm{LC}=\frac{\epsilon_0\Delta\epsilon E^2}%
{3a'_\mathrm{LC}(T-T^*-\Delta T_\mathrm{NI})}
\left[1+\frac{\phi_\mathrm{NP}}{3}
\left(\frac{4\pi P^2 R^3}{45\epsilon_0\epsilon k_\mathrm{B} T}\right)^2
\right] .
\end{equation}
In this expression, note that the induced order parameter depends on electric
field and temperature exactly as in the standard Kerr effect, but the
coefficient is increased by the coupling with nanoparticles.  In the square
brackets, the first term of 1 indicates the standard Kerr effect for pure
liquid crystals, and the second term indicates the relative enhancement due to
nanoparticle doping.

For a specific numerical example, we use the same parameters presented after
Eq.~(\ref{DeltaTNI}).  With these parameters, the relative enhancement in the
Kerr effect is extremely large, of order $10^7$.  Figure~3 plots the predicted
order parameter $S_\mathrm{LC}$ as a function of temperature for several
values of the applied electric field, with and without nanoparticles.  This
plot shows explicitly that the presence of nanoparticles greatly enhances the
sensitivity to applied electric fields in the isotropic phase, as well as
enhancing the isotropic-nematic transition temperature.  This prediction
should be tested in future experiments, and should provide an opportunity to
build liquid-crystal devices that can operate at lower electric fields.

\begin{figure}
\includegraphics[width=3.375in]{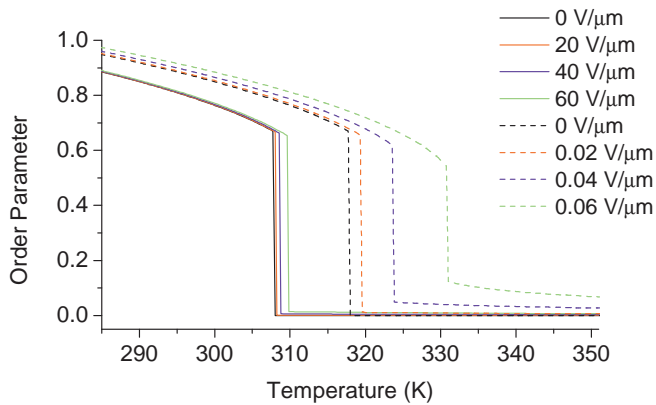}
\caption{(Color online) Prediction for field-induced order parameter
$S_\mathrm{LC}$ as a function of temperature for several values of applied
electric field, with and without ferroelectric nanoparticles, using
numerical parameters presented after Eq.~(\ref{DeltaTNI}).}
\end{figure}

As a final point, we should mention one limitation of our model.  Like all
Landau theories, our model involves an expansion of the free energy in powers
of the order parameters, and hence it overestimates the order parameters that
occur in the low-temperature phase.  Future work may extend this model through
asymptotic low-temperature approximations to the free energy.  Nevertheless,
our current model clearly shows that effects of nanoparticle doping at and
above the isotropic-nematic transition, in the regime where Landau theory is
valid.

In conclusion, we have developed a theory for the statistical mechanics of
ferroelectric nanoparticles in nematic liquid crystals.  This theory predicts
the enhancement in the isotropic-nematic transition temperature and in the
response to an applied electric field, which can be tested experimentally.
The work demonstrates the coupling of nanoparticles with macroscopic
orientational order, and provides an opportunity to improve the properties of
liquid crystals without chemical synthesis.

We would like to thank J. L. West, Y. Reznikov, and P. Bos for many helpful
discussions.  This work was supported by NSF Grant DMR-0605889.


\begin{thebibliography}{99}
\bibitem{reznikov} Y. Reznikov, personal communication.  Note that the
polarization of bulk Sn$_2$P$_2$S$_6$ is about 0.14 Cm$^{-2}$, but it is
reduced by surface effects in nanoparticles.
\end{thebibliography}
\end{document}